\documentstyle[12pt]{article}

\begin{document}

\newcommand \fr[2] {\mbox{${#1\over #2}$}}
\newcommand \FR[2] {\mbox{$\,\displaystyle{\strut\displaystyle #1 \over
 \strut\displaystyle #2}\,$}}
\newcommand \I[2] {\mbox{{\bf #1}$^{#2}$}}

\setlength\unitlength{1pt} 

\hyphenation {re-pre-sent-a-tion}

\thispagestyle{empty}
\setcounter{page}{0}

\begin{flushright}
JINR-E2-95-526
\end{flushright}

{}~\vfill

\begin{center}

{\Large\bf RECURRENCE RELATIONS \\*[10pt]
 FOR THREE-LOOP PROTOTYPES \\*[10pt]
 OF BUBBLE DIAGRAMS WITH A MASS}
 \footnote{~Supported in part by Volks\-wagen\-stiftung, RFFR grant
  \#\,94-02-03665, and JSPS FSU Project.}

\vfill

{\Large Leo. V. Avdeev}
 \footnote{~E-mail: $avdeevL@thsun1.jinr.dubna.su$ ~or~
  $avdeevL@hrz.uni$-$bielefeld.de$\,.}

\vfill

{\it Bogoliubov Laboratory of Theoretical Physics,\\
 Joint Institute for Nuclear Research,\\
 Dubna $($Moscow Region$)$, RU $141\,980$, Russian Federation}

\vfill

\begin{abstract}

Recurrence relations derived via the Chetyrkin--Tkachov method of
integration by parts are applied to reduce scalar three-loop bubble
(vacuum) diagrams with a mass to a limited number of master integrals.
The reduction is implemented as a package of computer programs for
analytic evaluation in FORM. The algorithms are applicable to diagrams
with any integer powers on the lines in an arbitrary dimension. A
physical application is the evaluation of the three-loop QCD correction
to the electroweak rho parameter.

\end{abstract}

\end{center}

{}~\vfill

\pagebreak

Vacuum (bubble) Feynman integrals (without external momenta) appear as
low-energy limits of certain physical amplitudes or as Taylor
coefficients of multipoint Green functions. The coefficients can then be
used to recover the functions in the whole complex plane of
momenta\,\cite{expand}. The presence of virtual heavy particles, like
the top quark in the quantum chromodynamics (QCD), generates an
effective high-energy scale, which makes perturbation theory applicable
owing to asymptotic freedom. In quadratically and linearly divergent
diagrams the contributions of heavy particles are enhanced by the power
of the mass, so that light particles can well be considered as massless.
This leaves us with an important special case of just one nonzero mass.

After performing the Dirac and Lorentz algebra, any vacuum diagram can
be reduced to some linear combinations of scalar bubble integrals. In
the three-loop case with the full tetrahedron topology of the diagram,
all scalar products in the numerator can be expressed through the
quadratic combinations in the denominators. Thus, in the most general
case, only a product of some powers of the denominators should be
integrated. A prototype defines the arrangement of massive and massless
lines in a diagram. Individual integrals are specified by the powers of
the denominators, called indices of the lines.

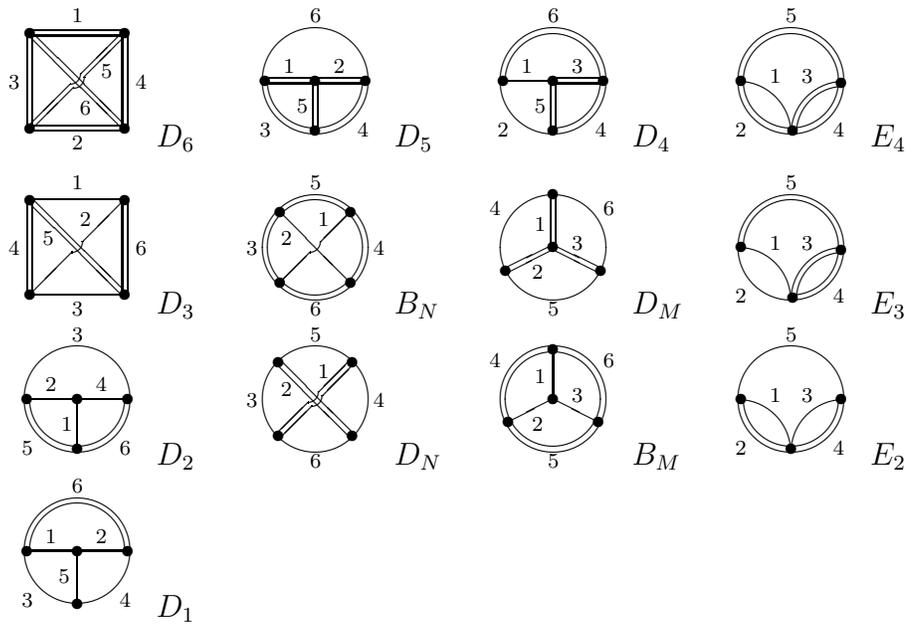
\begin{figure}[htbp]
 $$
 \begin{picture}(90,50)(-25,-25)
  \put(-19,-18){\line(0,1){36}}
  \put(-17,-18){\line(0,1){36}}
  \put(-18,-17){\line(1,0){36}}
  \put(-18,-19){\line(1,0){36}}
  \put(17,-18){\line(0,1){36}}
  \put(19,-18){\line(0,1){36}}
  \put(-18,17){\line(1,0){36}}
  \put(-18,19){\line(1,0){36}}
  \put(-18.7,17.3){\line(1,-1){36}}
  \put(-17.3,18.7){\line(1,-1){36}}
  \put(-2.7,2.7){\oval(8,8)[br]}
  \put(-1.3,1.3){\oval(8,8)[br]}
  \put(1.3,2.7){\line(1,1){16}}
  \put(2.7,1.3){\line(1,1){16}}
  \put(-1.3,-2.7){\line(-1,-1){16}}
  \put(-2.7,-1.3){\line(-1,-1){16}}
  \put(-18,18){\circle*4}
  \put(18,18){\circle*4}
  \put(-18,-18){\circle*4}
  \put(18,-18){\circle*4}
  \put(-2,22){\scriptsize 1}
  \put(-2,-26){\scriptsize 2}
  \put(-26,-3){\scriptsize 3}
  \put(22,-3){\scriptsize 4}
  \put(9,1){\scriptsize 5}
  \put(1,-13){\scriptsize 6}
  \put(30,-25){$D_6$}
 \end{picture}
 \begin{picture}(90,50)(-25,-25)
  \put(0,0){\circle{40}}
  \put(0,0){\oval(36,36)[b]}
  \put(-20,1){\line(1,0){40}}
  \put(-20,-1){\line(1,0){40}}
  \put(1,0){\line(0,-1){20}}
  \put(-1,0){\line(0,-1){20}}
  \put(0,0){\circle*4}
  \put(-19,0){\circle*4}
  \put(19,0){\circle*4}
  \put(0,-19){\circle*4}
  \put(-12,3){\scriptsize 1}
  \put(7,3){\scriptsize 2}
  \put(-21,-21){\scriptsize 3}
  \put(16,-21){\scriptsize 4}
  \put(-7,-12){\scriptsize 5}
  \put(-2,22){\scriptsize 6}
  \put(30,-25){$D_5$}
 \end{picture}
 \begin{picture}(90,50)(-25,-25)
  \put(0,0){\circle{40}}
  \put(0,0){\oval(36,36)[t]}
  \put(0,0){\oval(36,36)[br]}
  \put(-20,0){\line(1,0){20}}
  \put(0,1){\line(1,0){20}}
  \put(0,-1){\line(1,0){20}}
  \put(1,0){\line(0,-1){20}}
  \put(-1,0){\line(0,-1){20}}
  \put(0,0){\circle*4}
  \put(-19,0){\circle*4}
  \put(19,0){\circle*4}
  \put(0,-19){\circle*4}
  \put(-12,3){\scriptsize 1}
  \put(-21,-21){\scriptsize 2}
  \put(7,3){\scriptsize 3}
  \put(16,-21){\scriptsize 4}
  \put(-7,-12){\scriptsize 5}
  \put(-2,22){\scriptsize 6}
  \put(30,-25){$D_4$}
 \end{picture}
 \begin{picture}(90,50)(-25,-25)
  \put(0,0){\circle{40}}
  \put(0,0){\circle{36}}
  \put(-20,-20){\oval(40,40)[tr]}
  \put(20,-20){\oval(40,40)[tl]}
  \put(20,-20){\oval(36,36)[tl]}
  \put(-19,0){\circle*4}
  \put(19,-1){\circle*4}
  \put(.7,-19){\circle*4}
  \put(-8,-1){\scriptsize 1}
  \put(-21,-21){\scriptsize 2}
  \put(4,-1){\scriptsize 3}
  \put(16,-21){\scriptsize 4}
  \put(-2,22){\scriptsize 5}
  \put(30,-25){$E_4$}
 \end{picture} $$ $$
 \begin{picture}(90,50)(-25,-25)
  \put(-19,-18){\line(0,1){36}}
  \put(-17,-18){\line(0,1){36}}
  \put(-18,-18){\line(1,0){36}}
  \put(17,-18){\line(0,1){36}}
  \put(19,-18){\line(0,1){36}}
  \put(-18,18){\line(1,0){36}}
  \put(-18.7,17.3){\line(1,-1){36}}
  \put(-17.3,18.7){\line(1,-1){36}}
  \put(-2,2){\oval(8,8)[br]}
  \put(-2,-2){\line(-1,-1){16}}
  \put(2,2){\line(1,1){16}}
  \put(-18,18){\circle*4}
  \put(18,18){\circle*4}
  \put(-18,-18){\circle*4}
  \put(18,-18){\circle*4}
  \put(-2,22){\scriptsize 1}
  \put(1,8){\scriptsize 2}
  \put(-2,-26){\scriptsize 3}
  \put(-26,-3){\scriptsize 4}
  \put(-13,1){\scriptsize 5}
  \put(22,-3){\scriptsize 6}
  \put(30,-25){$D_3$}
 \end{picture}
 \begin{picture}(90,50)(-25,-25)
  \put(0,0){\circle{40}}
  \put(0,0){\circle{36}}
  \put(-13.4,13.4){\circle*4}
  \put(-13.4,-13.4){\circle*4}
  \put(13.4,-13.4){\circle*4}
  \put(13.4,13.4){\circle*4}
  \put(-13.4,13.4){\line(1,-1){26.8}}
  \put(-2,2){\oval(8,8)[br]}
  \put(-2,-2){\line(-1,-1){11.4}}
  \put(2,2){\line(1,1){11.4}}
  \put(1,8){\scriptsize 1}
  \put(-13,1){\scriptsize 2}
  \put(-26,-3){\scriptsize 3}
  \put(22,-3){\scriptsize 4}
  \put(-2,22){\scriptsize 5}
  \put(-2,-26){\scriptsize 6}
  \put(30,-25){$B_N$}
 \end{picture}
 \begin{picture}(90,50)(-25,-25)
  \put(0,0){\circle{40}}
  \put(0,0){\circle*4}
  \put(0,20){\circle*4}
  \put(-18,-9){\circle*4}
  \put(18,-9){\circle*4}
  \put(1,0){\line(0,1){20}}
  \put(-1,0){\line(0,1){20}}
  \put(0,1){\line(2,-1){18}}
  \put(0,-1){\line(2,-1){18}}
  \put(0,1){\line(-2,-1){18}}
  \put(0,-1){\line(-2,-1){18}}
  \put(-7,6){\scriptsize 1}
  \put(-8,-12){\scriptsize 2}
  \put(7,-1){\scriptsize 3}
  \put(-24,12){\scriptsize 4}
  \put(-2,-26){\scriptsize 5}
  \put(19,12){\scriptsize 6}
  \put(30,-25){$D_M$}
 \end{picture}
 \begin{picture}(90,50)(-25,-25)
  \put(0,0){\circle{40}}
  \put(0,0){\oval(36,36)[t]}
  \put(0,0){\oval(36,36)[br]}
  \put(-20,-20){\oval(40,40)[tr]}
  \put(20,-20){\oval(40,40)[tl]}
  \put(20,-20){\oval(36,36)[tl]}
  \put(-19,0){\circle*4}
  \put(19,-1){\circle*4}
  \put(.7,-19){\circle*4}
  \put(-8,-1){\scriptsize 1}
  \put(-21,-21){\scriptsize 2}
  \put(4,-1){\scriptsize 3}
  \put(16,-21){\scriptsize 4}
  \put(-2,22){\scriptsize 5}
  \put(30,-25){$E_3$}
 \end{picture} $$ $$
 \begin{picture}(90,50)(-25,-25)
  \put(0,0){\circle{40}}
  \put(0,0){\oval(36,36)[b]}
  \put(-19,0){\line(1,0){38}}
  \put(0,0){\line(0,-1){19}}
  \put(-19,0){\circle*4}
  \put(19,0){\circle*4}
  \put(0,-19){\circle*4}
  \put(0,0){\circle*4}
  \put(-6,-12){\scriptsize 1}
  \put(-12,3){\scriptsize 2}
  \put(-2,22){\scriptsize 3}
  \put(7,3){\scriptsize 4}
  \put(-21,-21){\scriptsize 5}
  \put(16,-21){\scriptsize 6}
  \put(30,-25){$D_2$}
 \end{picture}
 \begin{picture}(90,50)(-25,-25)
  \put(0,0){\circle{40}}
  \put(-14,14){\circle*4}
  \put(-14,-14){\circle*4}
  \put(14,-14){\circle*4}
  \put(14,14){\circle*4}
  \put(-14.7,13.3){\line(1,-1){28}}
  \put(-13.3,14.7){\line(1,-1){28}}
  \put(-2.7,2.7){\oval(8,8)[br]}
  \put(-1.3,1.3){\oval(8,8)[br]}
  \put(-2.7,-1.3){\line(-1,-1){12}}
  \put(-1.3,-2.7){\line(-1,-1){12}}
  \put(2.7,1.3){\line(1,1){12}}
  \put(1.3,2.7){\line(1,1){12}}
  \put(1,8){\scriptsize 1}
  \put(-13,1){\scriptsize 2}
  \put(-26,-3){\scriptsize 3}
  \put(22,-3){\scriptsize 4}
  \put(-2,22){\scriptsize 5}
  \put(-2,-26){\scriptsize 6}
  \put(30,-25){$D_N$}
 \end{picture}
 \begin{picture}(90,50)(-25,-25)
  \put(0,0){\circle{40}}
  \put(0,0){\circle{36}}
  \put(0,0){\circle*4}
  \put(0,19){\circle*4}
  \put(-17,-8.5){\circle*4}
  \put(17,-8.5){\circle*4}
  \put(0,0){\line(0,1){19}}
  \put(0,0){\line(2,-1){17}}
  \put(0,0){\line(-2,-1){17}}
  \put(-7,6){\scriptsize 1}
  \put(-8,-12){\scriptsize 2}
  \put(7,-1){\scriptsize 3}
  \put(-24,12){\scriptsize 4}
  \put(-2,-26){\scriptsize 5}
  \put(19,12){\scriptsize 6}
  \put(30,-25){$B_M$}
 \end{picture}
 \begin{picture}(90,50)(-25,-25)
  \put(0,0){\circle{40}}
  \put(0,0){\oval(36,36)[b]}
  \put(-20,-20){\oval(40,40)[tr]}
  \put(20,-20){\oval(40,40)[tl]}
  \put(-19,0){\circle*4}
  \put(19,0){\circle*4}
  \put(0,-19){\circle*4}
  \put(-8,-1){\scriptsize 1}
  \put(-21,-21){\scriptsize 2}
  \put(4,-1){\scriptsize 3}
  \put(16,-21){\scriptsize 4}
  \put(-2,22){\scriptsize 5}
  \put(30,-25){$E_2$}
 \end{picture} $$ $$
 \begin{picture}(90,50)(-25,-25)
  \put(0,0){\circle{40}}
  \put(0,0){\oval(36,36)[t]}
  \put(-20,0){\line(1,0){40}}
  \put(0,0){\line(0,-1){20}}
  \put(0,0){\circle*4}
  \put(-19,0){\circle*4}
  \put(19,0){\circle*4}
  \put(0,-20){\circle*4}
  \put(-12,3){\scriptsize 1}
  \put(7,3){\scriptsize 2}
  \put(-21,-21){\scriptsize 3}
  \put(16,-21){\scriptsize 4}
  \put(-7,-12){\scriptsize 5}
  \put(-2,22){\scriptsize 6}
  \put(30,-25){$D_1$}
 \end{picture} \hspace*{270pt} $$
 \caption{The three-loop scalar bubble prototypes with one mass. Double
  (single) lines refer to massive (massless) propagators in the momentum
  representation. Any line may have an integer index, a power of the
  denominator. The indices stay as arguments of the corresponding
  functions. Numbers define the ordering of the arguments.}
 \label{proto}
\end{figure}

In the dimensional regularization with $N=4-2\varepsilon$, any
massless bubbles are trivially equal to zero, so that at least one
massive line should be present. Fig.\,\ref{proto} displays all possible
three-loop prototypes. It was convenient also to distinguish some
reduced prototypes with a line missing, $E_{2-4}$, since they are
generated in evaluating several different full prototypes. The $B_M$ and
$B_N$ types have been completely analyzed in Ref.\,\cite{broad} and need
not be discussed here.

The method of recurrence relations\,\cite{parts,broad} connects
integrals of the same prototype but with different values of the
indices. Using these relations ingeniously enough, one can reduce any
integral to a limited set of so-called master integrals. However, that
remains still a kind of art without any strict assertions as to the
minimal set of the master integrals or the most efficient strategy. Let
us derive a relation\,\cite{kitty} for a generic triangle subgraph with
masses on its lines $m_1$, $m_2$, and $m_3$, line momenta $p_1$,
$p_2=p_1-p_{12}$, and $p_3=p_1-p_{13}$, and indices $j_1$,
$j_2$, and $j_3$, respectively:
\begin{eqnarray}
 &\displaystyle 0 = \int {\rm d}^N p_1 \FR \partial {\partial p_1^\mu}
  \FR {p_1^\mu} { c_1^{j_1} c_2^{j_2} c_3^{j_3} }
  = \int \FR {{\rm d}^N p_1} { c_1^{j_1} c_2^{j_2} c_3^{j_3} }
  \Big( N -2 j_1 -j_2 -j_3 +j_1 \FR {2 m_1^2} {c_1}& \nonumber \\*
 & +\,j_2 \FR {m_1^2 +m_2^2 -m_{12}^2 +c_{12} -c_1} {c_2}
   +j_3 \FR {m_1^2 +m_3^2 -m_{13}^2 +c_{13} -c_1} {c_3}
  \Big) \, ,& \label{123}
\end{eqnarray}
where $c_k=p_k^2+m_k^2$. Dividing or multiplying by $c_k$ just
increments or decrements index $j_k$ of the $k\,$th line. The
corresponding operator can be denoted by \I{K}{\pm} \cite{broad}. For an
arbitrary $L$-loop diagram, integrating by parts the first derivatives
provides us with $L^2$ linearly independent relations in total. It is
convenient to refer to triangle recurrence relations by specifying only
the line numbers: \{123\} for Eq.\,(\ref{123}), line\,1 being the base.
Half sum of the relations for three faces of a tetrahedron as their base
lines form a triangle, like \{124\}, \{534\}, and \{623\} in $D_3$,
gives a relation \{dim\} that is evident on dimensional grounds:
\begin{equation}
 \big[ \fr 3 2\,N -j_1-...-j_6 +m^2 (j_4\,\I4+ +j_5\,\I5+ +j_6\,\I6+)
 \big] D_3 (j_1,...,j_6) = 0\, . \label{dim}
\end{equation}

Now follows a brief description of evaluating various prototypes. In
Fig.\,\ref{proto} they have been ordered from the `most difficult' to
the `simplest'. Whenever in $D_6$ a denominator is absent ($j_k\le 0$)
the diagram is immediately reduced to the `simpler' $D_5$ type by
expanding the power of the polynomial. A typical trick to bring any
positive index down to 1 is as follows. A combination is sought, like
$3\{146\}-\{416\}-\{614\}$ for $j_1>1$ in $D_6$, in which only one
`highest' term $m^2 j_1 \I1+$ is present, all others having the sum of
the indices less by one. The highest term can be expressed through the
others until the index on the line reaches~1. Thus we arrive at the
master integral $D_6(1,...,1)$.

If $D_5$ has $j_5\le 0$, it is reduced to $B_N$; $j_{1-4}\le 0$ to
$D_4$. As $j_6>0$, it is profitable to solve \{612\} (no $m^2$ for a
massless exchange between particles of unchanged masses\,\cite{kitty})
with respect to the free term. Eventually, this brings $j_{3,4}$ or
$j_6$ to zero. To get rid of the numerator $j_6<0$, a denominator on
an adjacent line is typically used: if for example $j_1>1$, \{315\}
can be solved relative to \I1+\,\I6-. Otherwise, as $j_{1-4}=1$, the
quadratic denominators can be created by solving \{126\} with respect to
the free term. The irreducible master integral is $D_5(1,1,1,1,1,0)$.

In $D_4$, $j_6\le 0$ leads to $B_M$; $j_5\le 0$ to $D_M$;
$j_{3,4}\le 0$ to $D_3$; and $j_{1,2}=0$ to $E_4$. The numerator on
line\,1 can be eliminated by \{246\} if $j_6>1$; by 2\{dim\}$-$\{534\}
if $j_3>1$; by \{dim\}$-$\{345\} if $j_5>1$; by \{dim\}$-$\{624\} if
$j_2\ne 1$; or by \{215\} otherwise. Solving \{215\} relative to $m^2
\I1+$ diminishes the denominator $j_1>1$; \{512\} reduces $j_5>1$;
\{435\}$-$\{534\} reduces $j_3>1$; \{136\}+\{246\} reduces $j_6>1$,
leading to $D_4(1,...,1)$.

As $j_{4,6}\le 0$ in $D_3$, this is $D_2$; $j_5\le 0$ $\Rightarrow$
$D_N$; $j_2\le 0$ $\Rightarrow$ $B_N$. The numerator $j_1<0$ is normally
reduced by 2\{dim\}$-$\{534\} as $j_4>1$; by \{623\} as $j_2>1$; by
\{236\} as $j_6>1$; by 2\{dim\}$-$\{435\} as $j_5>1$; and \{dim\}
creates $j_{4-6}>1$ otherwise. The special case $j_1=j_3=0$ is more
efficiently worked out as $E_4$ with $j_5=0$. The remaining denominator
on line\,5 in $D_3$ is brought down to 1 by \{516\}; $j_2>1$ by
$(1-j_2)\{236\} +j_3\I3+\I2-\{326\} +j_6\I6+\I2-\{623\}$; and $j_6>1$
by $\{156\}-\{534\} +(2+\I1-/m^2)$\{dim\}. However, that may revive
massless numerators $j_{1,3}<0$. To avoid infinite loops on recursive
application of the relations, we eliminate single numerators by a
general projection-operator method\,\cite{parts}:
\begin{eqnarray}
 &\displaystyle \int {\rm d}^N p_1 ~ {\rm d}^N p_2 ~
  f_1[p_1^2,(p_1-q)^2]~ f_2[p_2^2,(p_2-q)^2]~ A^{2n}(p_1,p_2,q) \,=&
  \nonumber\\*
 &\displaystyle
  \FR { \Gamma(n+\fr 1 2)\, \Gamma \big[ \fr 1 2 (N-1) \big] }
   { \Gamma(\fr 1 2)\, \Gamma \big[ n +\fr 1 2 (N-1) \big] }
  \prod_{j=1}^2 \int {\rm d}^N p_j~ f_j[p_j^2,(p_j-q)^2]~ A^n(p_j,p_j,q)
  \, ,& \label{int}
\end{eqnarray}
where $A(p_1,p_2,q) = 4\, p_1^\mu \, \big( g_{\mu\nu} -q_\mu q_\nu /q^2
\big) \, p_2^\nu\,$, and $f_{1,2}$ are arbitrary functions of their
scalar arguments. A numerator $\big[(p_1-p_2)^2\big]^n$ can be
re-expanded:
\begin{eqnarray}
 &(p_1-p_2)^2 = \fr 1 2 \Big\{ p_1^2 +(p_1-q)^2 +p_2^2 +(p_2-q)^2&
  \nonumber\\*
 &-\, \big[ p_1^2 -(p_1-q)^2 \big] \big[ p_2^2 -(p_2-q)^2 \big] /q^2
  -q^2 -A(p_1,p_2,q) \Big\} .& \label{num1}
\end{eqnarray}
Odd powers of $A(p_1,p_2,q)$ fall out after integration, and for even
powers Eq.\,(\ref{int}) yields $A(p,p,q) = 2 \big[ p^2 +(p-q)^2 \big] -
\big[ p^2 -(p-q)^2 \big]^2 /q^2 -q^2$. For $D_3$ we identify the
left-hand side of Eq.\,(\ref{num1}) with $c_1$, and $q^2$ with $c_3$. If
we deal with $j_1=-1$, $j_3=0$, $j_4=j_6$, then $q^2$ on the
right-hand side of Eq.\,(\ref{num1}) generates the original integral
which can then be eliminated.

The denominator $j_3>1$ is reduced by \{156\}$-$\{236\} while
$j_1<0$. At $j_1=0$ with $j_{2,4-6}=1$ we apply \{236\}, eliminate
$j_1=-1$ by Eqs.\,(\ref{int}) and (\ref{num1}), reduce $j_6=2$ as
usual, and expand the first numerator again. The original integral with
the same value of $j_3$ can be expressed by solving the resulting
equation.

The case $D_3(0,1,1,1,1,1)$ can be transformed as follows. The
differential equation for the two-loop subgraph without line\,2,
Eq.\,(43) of Ref.\,\cite{pro2}, is divided by $q^2(q^2+m^2)$ and
integrated over~$q$. Derivatives with respect to $m^2$ are taken via
\{dim\}, and after substituting simple integrals we get
\begin{eqnarray}
 &D_3(0,1,1,1,1,1) = \FR {3N-8} {2 m^2} D_3(0,1,0,1,1,1)& \nonumber\\*
 &+~ \FR {8 \left( m^2 \right)^{3N/2-5}} {(N-2)(N-3)(N-4)^3}
  \Big[ \FR {\Gamma(\fr 1 2 N-1)~ \Gamma(5-N)} {\Gamma(3-\fr 1 2 N)} +4
  \Big] \, ,&
\end{eqnarray}
where each loop integral was divided by $\pi^{N/2}\Gamma(3-\fr 1 2 N)$;
in one loop, this modification agrees with the standard $\overline{\rm
MS}$ definition but is more convenient in higher-loop massive
calculations. As a result of the transformations, any $D_3$-type
diagram is reduced to simpler types and two master integrals
$D_3(0,1,0,1,1,1)$ and $D_3(1,...,1)$.

With $j_{3,4}\le 0$, $E_4$ is a particular case of $B_M$. The numerator
on line\,5 can be taken off by \{120\} as $j_2>1$ (massless line\,0
with index\,0 is assumed to connect 1.3 and 2.4); by \{210\} as
$j_1\ne 1$; by \{430\} as $j_3>1$; or \{dim\} should be applied
otherwise. The $j_1<0$ numerator is eventually reduced to $j_1=0$
(hence, to the two-loop massive bubbles) by \{120\} if $j_2>1$, or by
\{210\}. The denominators $j_{1,2}>1$ can be brought down to 1 by
solving \{120\} relative to \I2+\I5-, or $\{210\}-2m^2\I5+\{120\}$
relative to \I1+\I5-. That always increases $j_5$. The latter helps to
reduce the denominators in the massive one-loop subgraph by
2\{340\}$-$\{430\}. The extra denominator $j_5>0$ can be integrated
off by applying \{dim\} and reducing $j_2=2$ by \{120\}. Iteratively,
we arrive at the master integral $E_4(1,1,1,1,0)=D_3(0,1,0,1,1,1)$.

Further prototypes are quite simple indeed. The numerators are
manageable by ad\-ja\-cent-tri\-angle relations, and the denominators
can be reduced by combining three relations for a triangle that
contains the line. The master integrals are $D_M(1,...,1)$,
$E_3(1,...,1)$, and $D_N(1,...,1)$. In $D_2$ the massless relation
\{125\} solved with respect to the free term allows one to cancel out a
denominator. Thus, in the end $D_2$ is reduced to $\Gamma$ functions,
just as $D_1$ does.  For $E_2$ a massless relation can be constructed
as \{524\}$-$\{dim\}.

The described algorithms are implemented as a package of procedures in
the symbolic-manipulation language FORM\,\cite{form} well suited to
evaluating the Feynman diagrams as well as any polynomial-like
expressions with a large number of terms. However, the efficiency of the
essentially recursive programs is restricted by some features of the
existing FORM translator. In particular, `infinitely' iterative
substitutions are only allowed without any intermediate sorting of
terms. On the other hand, the recurrence relations generate rather many
equal terms, and after exceeding certain machine-dependent limits on the
size of the scratch expression generated in a module, the sorting
becomes extremely slow. The only way out is to use step-by-step sorting
inside the preprocessor {\bf \#do} loops with a pre-estimated number of
repetitions. But sometimes the number is rather difficult to guess at
beforehand, while any misjudgement spoils the program performance.

Therefore, it would be highly desirable to implement a kind of the
preprocessor {\bf \#repeat/endrepeat} construct into FORM, which would
terminate as soon as no actual transformations are applicable in any
module inside its body\,\footnote{~I thank Timo van Ritbergen for
informing me that in an wxperimental version of FORM 2.2 it is possible
to terminate {\bf \#do} loops as nothing changes. However, this
undocumented feature is unavailavle in public versions and, as Jos A.M.
Vermaseren communicates, liable to changes in FORM~3}. Also of use
would be any means of redefining preprocessor variables, based on
global tests on the terms of sorted expressions. An invariable essential
inconvenience for structured packages is the global scope of all names
in FORM.

The first application of the described package was the evaluation of the
three-loop QCD correction to the electroweak $\rho$
parameter\,\cite{rho}:
\begin{eqnarray}
&\delta^{QCD} = -\fr 2 3 \big[ 1+2\,\zeta(2) \big] \FR {\alpha_s} {\pi}
 + \Big\{ \fr{157}{648} -\fr{3313}{162}\,\zeta(2)
 -\fr{308}{27}\,\zeta(3) +\fr{143}{18}\,\zeta(4)& \nonumber \\*
&-\,\fr 4 3\,\zeta(2) \ln 2 +\fr{441}{8}\,S_2 -\fr 1 9\,B_4
 -\fr 1{18}\,D_3
 -\big[ \fr 1{18} -\fr{13}9\,\zeta(2) +\fr 4 9\,\zeta(3) \big] n_f&
 \nonumber \\*
&-\, \big( \fr{11}{6} -\fr 1 9\,n_f \big) \big[ 1+2\,\zeta(2) \big]
 \ln\,(\mu^2/m_t^2) \Big\} \Big( \FR {\alpha_s} {\pi} \Big)^2,&
 \label{rho3}
\end{eqnarray}
where $\alpha_s$ is the QCD coupling constant at the renormalization
scale $\mu$ in the $\overline{\rm MS}$ scheme with the total number of
quark flavors $n_f$ (=6); $m_t$ is the pole mass of the top quark; $B_4$
has been introduced in Ref.\,\cite{broad}; $S_2$ determines the finite
part of the two-loop massive bubble master integral\,\cite{pro2}; and
$D_3$ is the finite part of $D_3(1,1,1,1,1,1)$ which has been evaluated
numerically by the momentum-expansion method\,\cite{expand} and
independently in Ref.\,\cite{check}. An error in the coefficient of
$\zeta(4)$ in the original publication has been fixed, so that
Eq.\,(\ref{rho3}) completely agrees with the independent calculation of
Ref.\,\cite{check}.
\vspace*{20pt}

\noindent
{\Large\bf Acknowledgments}\\*[10pt]
I am grateful to the Physics Department of the University of Bielefeld
where a part of this work was done. The financial support of
Volks\-wagen\-stiftung, RFFR grant \#\,94-02-03665, and JSPS FSU
Project is thankfully acknowledged. Scientific discussions with O.V.
Tarasov were especially useful.
\pagebreak[4]

\end{document}